\documentclass[twocolumn,groupedaddress,amsmath,amssymb,prb,aps]{revtex4-2}

\usepackage{graphicx}
\usepackage{dcolumn}
\usepackage{bm}
\usepackage{soul} 
\usepackage{xcolor} 
\usepackage{amsmath} 

\begin{document}

\title{Coexisting conventional and inverse mechanocaloric effects in ferroelectrics}

\author{D.E. Murillo-Navarro$^{1,2}$, M\'onica Graf$^{1}$ and Jorge \'I\~niguez$^{1,2}$}

\affiliation{
  \mbox{$^{1}$Materials Research and Technology Department,
    Luxembourg Institute of Science and Technology (LIST),} \mbox{Avenue
    des Hauts-Fourneaux 5, L-4362 Esch/Alzette,
    Luxembourg}\\
 \mbox{$^{2}$Department of Physics and Materials Science, University
   of Luxembourg, Rue du Brill 41, L-4422 Belvaux, Luxembourg} }
   
\begin{abstract}
The mechanocaloric effect is the temperature change of a material upon
application or removal of an external stress. Beyond its fundamental
interest, this caloric response represents a promising and ecofriendly
alternative to current cooling technologies. To obtain large
mechanocaloric effects, we need materials whose elastic properties
(e.g., strain, elastic compliance) are strongly temperature
dependent. This is the case of ferroelectric perovskite oxides, where
the development of the spontaneous electric polarization is
accompanied by significant strains and lattice softening. Thus, in
this work we study the mechanocaloric properties of model
ferroelectric PbTiO$_{3}$, by means of predictive atomistic
(``second-principles'') simulations and a perturbative formalism here
introduced. Our calculations reveal relatively large effects (up to
$-$4~K for relatively small applied compressions of $-$0.1~GPa) and
several striking features. In particular, we find that the
mechanocaloric response is highly anisotropic in the ferroelectric
phase, as it can be either conventional (temperature increases upon
compression) or inverse (temperature decreases) depending on the
direction of the applied stress. We discuss and explain these
surprising results, which compare well with existing experimental
information. Our analysis suggests that the coexistence of
conventional and inverse mechanocaloric responses is probably common
among ferroelectrics and materials displaying a negative thermal
expansion.
\end{abstract}

\maketitle

\section{Introduction}

A material subject to an external stress changes its temperature
and/or entropy by virtue of the so-called mechanocaloric effect (also
denoted elastocaloric or barocaloric, respectively, when the applied
stress is uniaxial or
hydrostatic)~\cite{engelbrecht19,kabirifar19,imran21,chauhan15}. Mechanocaloric
effects in solid-state systems can be large, reaching values from 5~K
to 40~K in shape-memory alloys~\cite{manosa10}, super-ionic
conductors~\cite{aznar17}, or plastic crystals~\cite{lloveras19}. This
remarkable performance makes mechanocalorics a viable alternative for
applications in solid-state cooling, where it is pressing to replace
current technologies (e.g., compressed gas) that are polluting, noisy,
and difficult to downscale.

Mechanocaloric effects are large in ferroelastic materials, which are
strongly responsive to mechanical perturbations and often display
stress-driven phase transformations, particularly close to the
ferroelastic transition point~\cite{salje-book1993}. Interestingly,
ferroelectrics (e.g., perovskite oxides like BaTiO$_{3}$ and
PbTiO$_{3}$) are known to be very sensitive to mechanical
perturbations too~\cite{lines-book1977}, and can in principle be
expected to present a considerable mechanocaloric
response. Nevertheless, while ferroelectrics have been thoroughly
studied in regards to electrocaloric properties~\cite{kutnjak15},
their mechanocaloric performance has been scarcely investigated so
far~\cite{mikhaleva12,lisenkov13,barr15,khassaf17}.

In this article we present a theoretical study of the intrinsic
mechanocaloric response of prototype ferroelectric perovskite
PbTiO$_{3}$. Our work is based on predictive atomistic
(``second-principles'') simulations
\cite{wojdel13,escorihuelasayalero17,garciafernandez16} of the
temperature-dependent elastic properties of PbTiO$_{3}$, combined with
a perturbative formalism that we introduce here and should be useful
in the broader context of mechanocaloric investigations, including
experimental ones. Among other remarkable features, we find that the
ferroelectric state of PbTiO$_{3}$ presents a acute mechanocaloric
anisotropy, to the extent that the sign of the adiabatic temperature
change depends on the direction along which the stress is applied. We
argue that this feature is probably common among ferroelectric
compounds, as well as in materials displaying negative thermal
expansion.

\section{Methods}

In the following we describe our perturbative formalism to compute and
interpret the mechanocaloric response, as well as the
second-principles simulation methods that allow us to study the
specific case of PbTiO$_{3}$.

\subsection{Perturbative formalism}

We adapt to the mechanocaloric case a perturbative formalism recently
introduced by some of us in the context of electrocaloric
effects~\cite{graf21}. Our starting point is the usual thermodynamic
expression for the adiabatic temperature change caused by an applied
stress~\cite{chauhan15},
\begin{equation}
\begin{split}
	\Delta T(\sigma_a) & = -\int_{0}^{\sigma_a} {\frac{T}{C_\sigma}}\left(\frac{\partial \eta_a}{\partial T}\right)_\sigma d\sigma'_a\\
	& = -\int_{0}^{\sigma_a} {\frac{T}{C_\sigma}} \alpha_{a}(T,\sigma'_{a}) d\sigma'_a \; ,
	\end{split} 
\label{eq:general-deltaT}
\end{equation}
where the integral runs from zero to the final stress $\sigma_{a}$,
$\eta_a$ denotes the strain conjugate to the applied stress, $T$ the
temperature and $C_{\sigma}$ the specific heat at constant
stress. Here we introduce the thermal expansion coefficient
$\alpha_{a}$, indicating that it depends on temperature and
stress. The subscript $a$ labels stress and strain components in Voigt
notation \cite{nye-book1957} and we assume a Cartesian reference. Note
that we write $\Delta T(\sigma_{a})$ to indicate explicitly that the
temperature change can be anisotropic, i.e., it may depend on the
direction of the applied stress. Finally, we adopt the usual sign
convention that a negative external stress corresponds to a
compression of the material.

As usually done in the literature~\cite{kabirifar19}, in the following
we work with an approximate version of the temperature change,
\begin{equation}
	\Delta T(\sigma_a) \approx
        -\frac{T^{(0)}}{C_\sigma^{(0)}}\int_{0}^{\sigma_a}
        {\alpha_a\left(T^{(0)},\sigma'_a\right) d\sigma'_a} \; ,
\label{eq:approx-deltaT}
\end{equation}
where the ``0'' superscripts indicate values calculated at zero
applied stress and we assume that $C_{\sigma}$ and $\alpha_{a}$ are
independent from the mechanocaloric temperature change. (This is a
reasonable approximation since, in most situations of interest,
$|\Delta T(\sigma_{a})|\ll T^{(0)}\approx 300$~K.) Besides being
simpler and easier to treat, this formula allows us to focus on how
the thermal expansion $\alpha_{a}$ controls the effect. Note that this
is the quantity that can be expected to be most strongly $T$- and
stress-dependent; and the only one whose sign is undetermined and free
to change.

Now we write the strain in terms of the stress as a Taylor series,
\begin{equation}
	\eta_a = \eta_a^{(0)} + \sum_{b}S_{ab}^{(0)} \sigma_b + ... \; ,
\label{eq:eta}
\end{equation}
where $\eta_a^{(0)}$ stands for the spontaneous strain at zero stress;
$S_{ab}^{(0)}$ is the compliance tensor at zero stress, with
\begin{equation}
S_{ab} = \frac{\partial \eta_a}{\partial \sigma_b} \; .
\end{equation}
Taking the temperature derivative of Eq.~(\ref{eq:eta}), we obtain
\begin{equation}
	\alpha_a = \alpha_a^{(0)} + \sum_{b} \alpha_{ab}^{(0)} \sigma_b+... \; ,
\label{eq:alpha}
\end{equation}
where $\alpha_a^{(0)}$ is the thermal expansion vector of the
unperturbed material and we have
\begin{equation}
	\alpha_{ab}^{(0)}=\frac{\partial S_{ab}^{(0)}}{\partial T}\biggr|_{\sigma} \; ,
\end{equation}
which we can view as stress-induced thermal expansion
coefficients. Using these simple expressions, now we evaluate
Eq.~(\ref{eq:approx-deltaT}) in several cases.

\subsubsection{Uniaxial stress}

Imagine we apply stress along a specific Cartesian axis $a$, so that
$\sigma_b = \delta_{ab}\sigma_a$ where $\delta_{ab}$ is the Kronecker
delta. We can then substitute Eq.~(\ref{eq:alpha}) into
Eq.~(\ref{eq:approx-deltaT}) and solve the integral to obtain
\begin{equation}
\begin{split}
	\Delta T(\sigma_a) & =
        -\frac{T^{(0)}}{C_\sigma^{(0)}}\left(\alpha_a^{(0)}\sigma_a+\frac{1}{2}\alpha_{aa}^{(0)}\sigma_a^2+...\right)\\ &=\Delta
        T^{(1)}(\sigma_a) +\Delta T^{(2)}(\sigma_a) + ... \; ,
\end{split}
\label{eq:pert-deltaT}
\end{equation}
where we introduce $\Delta T^{(n)}(\sigma_a)$ to denote the
contribution of $n$-th order to the total adiabatic temperature
change.

\subsubsection{Biaxial stress}

Suppose now that we apply stress along two Cartesian directions
simultaneously. To fix ideas, let us imagine we work with $\sigma_{1}$
and $\sigma_{2}$, both varying from zero to a final value $\sigma$,
the generalization of our arguments being trivial.

When we work with two stress components, the one-dimensional integrals
presented above are not directly applicable, and we need to take some
extra intermediate steps. Let us discuss two equivalent ways to
compute the mechanocaloric temperature change in this case.

\paragraph*{Method~I.} We can rotate our Cartesian coordinates so that one of the axes coincides with the direction of the applied stress, and thus recover a one-dimensional problem. In this particular example, we can use the unitary transformation
\begin{equation}
    \begin{pmatrix}
    \sigma_{\parallel} \\ \sigma_{\perp}
    \end{pmatrix}
    =\frac{1}{\sqrt{2}}
    \begin{pmatrix}
    1 & 1 \\
    1 & -1
    \end{pmatrix}
    \begin{pmatrix}
    \sigma_{1}\\
    \sigma_{2}
    \end{pmatrix} \; ,
\end{equation}
where $\sigma_{\parallel} = 2^{-1/2}(\sigma_{1}+\sigma_{2})$ defines
the stress we intend to apply. (The inverse transformation gives
$\sigma_{1}=\sigma_{2}=2^{-1/2}\sigma_{\parallel}$, assuming
$\sigma_{\perp}=0$.) The strain $\eta_{\parallel}$, conjugate to
$\sigma_{\parallel}$, is similarly given by $\eta_{\parallel} =
2^{-1/2}(\eta_{1}+\eta_{2})$, and we also have $\alpha_{\parallel} =
2^{-1/2}(\alpha_{1} + \alpha_{2})$ for the corresponding thermal
expansion coefficient. We have thus defined an equivalent
one-dimensional problem where the applied stress $\sigma_{\parallel}$
varies from zero to $\sqrt{2}\sigma$.

Now, to compute the temperature change, we need to solve the integral
\begin{equation}
	\Delta T(\sigma_{\parallel}) \approx -\frac{T^{(0)}}{C_\sigma^{(0)}}\int_{0}^{\sqrt{2}\sigma} {\alpha_{\parallel}\left(T^{(0)},\sigma'_{\parallel}\right) d\sigma'_{\parallel}} \; ,
\end{equation}
where
\begin{equation}
    \alpha_{\parallel} =
    \frac{1}{\sqrt{2}}(\alpha^{(0)}_{1}+\alpha^{(0)}_{2}) +
    \frac{1}{2}
    (\alpha^{(0)}_{11}+\alpha^{(0)}_{22}+2\alpha^{(0)}_{12})\sigma'_{\parallel}
    \; .
\end{equation}
Here we have used the relation $\alpha^{(0)}_{12} =
\alpha^{(0)}_{21}$, which relies on the fact that the compliance
matrix $\mathbf{S}$ is symmetric. Also, for conciseness, we have
truncated the series to the linear order in the applied stress. We can
finally evaluate the integral to obtain
\begin{equation}
\begin{split}
    \Delta T(\sigma_{\parallel}) = & \;
    -\frac{T^{(0)}}{C_\sigma^{(0)}} \left(
    (\alpha^{(0)}_{1}+\alpha^{(0)}_{2})\sigma \right. \\ & \; \left. +
    \frac{1}{2}(\alpha^{(0)}_{11}+\alpha^{(0)}_{22}+2\alpha^{(0)}_{12})\sigma^2
    \right) \; .
\end{split}
\label{eq:deltaT-methodI}
\end{equation}

\paragraph*{Method~II.} Alternatively, we can imagine that the stress is applied in two
steps. (The result of our integral for the temperature change does not
depend on how we apply the stress, a property that relies on the fact
that the underlying free energy is (assumed to be) an exact
differential.)

In the first step, we have $\sigma_{1}$ varying from zero to $\sigma$,
with $\sigma_2=0$ throughout. The corresponding temperature change is
given by the original one-dimensional integral
\begin{equation}
\begin{split}
        \left. \Delta T(\sigma_{1})\right|_{\sigma_{2}=0} & =
        -\frac{T^{(0)}}{C_\sigma^{(0)}} \int_{0}^{\sigma}
        \left.\alpha_{1}(\sigma'_{1})\right|_{\sigma_{2}=0}
        d\sigma'_{1} \\ & = -\frac{T^{(0)}}{C_\sigma^{(0)}} \left(
        \alpha^{(0)}_{1}\sigma + \frac{1}{2}\alpha_{11}^{(0)} \sigma^{2}
        \right) \; .
        \end{split}
\end{equation}
In the second step, we have a constant $\sigma_{1}=\sigma$, while
$\sigma_{2}$ varies from zero to $\sigma$. The corresponding
one-dimensional integral is
\begin{equation}
        \left. \Delta T(\sigma_{2})\right|_{\sigma_{1}=\sigma} = -\frac{T^{(0)}}{C_\sigma^{(0)}}  \int_{0}^{\sigma} \left.\alpha_{2}(\sigma'_{2})\right|_{\sigma_{1}=\sigma} d\sigma'_{2} \; ,
        \label{eq:deltaT-2nd-step}
\end{equation}
where one must note that 
\begin{equation}
    \left.\alpha_{2}(\sigma'_{2})\right|_{\sigma_{1}=\sigma} = \alpha^{(0)}_{2} + \alpha^{(0)}_{22}\sigma'_{2} + \alpha^{(0)}_{12}\sigma \; .
    \label{eq:alpha-2nd-step}
\end{equation}
Hence, inserting Eq.~(\ref{eq:alpha-2nd-step}) into Eq.~(\ref{eq:deltaT-2nd-step}), we get

\begin{equation}
    \left. \Delta T(\sigma_{2})\right|_{\sigma_{1}=\sigma} = -\frac{T^{(0)}}{C_\sigma^{(0)}} \left( \alpha^{(0)}_{2}\sigma + \frac{1}{2}\alpha^{(0)}_{22}\sigma^2 + \alpha^{(0)}_{12}\sigma^{2} \right) \; . 
\end{equation}

Finally, we calculate the total temperature change by adding the individual variations, to obtain
\begin{equation}
\begin{split}
    \Delta T (\sigma_{1}=\sigma_{2}) = & \;
    -\frac{T^{(0)}}{C_\sigma^{(0)}} \left( (\alpha^{(0)}_{1} +
    \alpha^{(0)}_{2})\sigma \right. \\ & \left. + \frac{1}{2}
    (\alpha^{(0)}_{11}+\alpha^{(0)}_{22}+2\alpha^{(0)}_{12}) \sigma^{2}
    \right) \; ,
    \end{split}
\end{equation}
which coincides exactly with the outcome of Method~I described above
(Eq.~(\ref{eq:deltaT-methodI})).

It is worth noting that this two-step approach is, in essence,
directly applicable to any multicaloric effect, where, instead of
considering two different components of the same field (stress in our
case), one applies two fields of different nature (e.g., mechanical
and electric).

\subsubsection{Hydrostatic pressure}

We finish with the case of an isotropic stress (hydrostatic pressure)
where all $\sigma_{1}=\sigma_{2}=\sigma_{3}$ vary from zero to
$\sigma$. The two methods described above for the biaxial case are
readily applicable, the final result for the adiabatic temperature
change being
\begin{widetext}
\begin{equation}
    \Delta T (\sigma_{1}=\sigma_{2}=\sigma_{3}) = -\frac{T^{(0)}}{C_\sigma^{(0)}} \left( (\alpha^{(0)}_{1} + \alpha^{(0)}_{2} + \alpha^{(0)}_{3})\sigma 
    + \frac{1}{2} (\alpha^{(0)}_{11}+\alpha^{(0)}_{22}+\alpha^{(0)}_{33} + 2\alpha^{(0)}_{12} + 2\alpha^{(0)}_{23} + 2\alpha^{(0)}_{31}) \sigma^{2} \right) \; .
\end{equation}
\end{widetext}

\subsection{Simulations for PbTiO$_{3}$}

In this work we evaluate the quantities controlling the mechanocaloric
temperature change (specific heat, spontaneous thermal expansion,
etc.) by running Monte Carlo simulations of an atomistic
second-principles model of PbTiO$_{3}$. This model, described in
detail in Ref.~\onlinecite{wojdel13}, has successfully been used in
several theoretical investigations of PbTiO$_{3}$ and related
compounds~\cite{wojdel14a,zubko16,goncalves19}, including a recent
study of the electrocaloric response~\cite{graf21}. Its only
noteworthy deficiency is that it predicts a ferroelectric transition
temperature that is too low as compared with the experimental one
(510~K {\it vs} 760~K). However, as we show below, our results for the
key quantities of interest here (most notably, the thermal expansion
and specific heat) compare very well with experimental values upon a
simple temperature shift that makes the theoretical and experimental
Curie points coincide; hence, this quantitative discrepancy is not
critical for the present purposes.

In the simulations we typically use a periodically-repeated supercell
containing 10$\times$10$\times$10 instances of the 5-atom perovskite
unit cell; we run 10,000 Monte Carlo sweeps for thermalization and
100,000 more to compute averages. Near the ferroelectric transition
temperature we find it necessary employ a larger
12$\times$12$\times$12 supercell, to reduce finite-size effects; in
that case we run 10,000 thermalization sweeps followed by 75,000
production sweeps. For the sake of simplicity, in the results
presented below the ferroelectric polarization is always chosen to lie
along the $z$ axis.

To obtain thermodynamic properties (e.g., equilibrium strain, specific
heat, elastic compliance tensor) from our Monte Carlo simulations, we
proceed in the usual manner, and rely on well-known linear-response
formulas (see, e.g., the description in Ref.~\cite{graf21}). To
illustrate the linear-response approach, let us consider the case of
the elastic compliance, which we have not been able to find in
previous literature. The compliance component $S_{ab}$ is calculated
as
\begin{equation}
    S_{ab} = \frac{\partial \langle \eta_{a} \rangle}{\partial \sigma_{b}} \; ,
\end{equation}
where $\langle \eta_{a} \rangle$ is the average equilibrium value of
the strain at given conditions of temperature and applied stress. More
specifically, $\langle \eta_{a} \rangle$ is
\begin{equation}
    \langle \eta_{a} \rangle = \frac{1}{Z} \sum_{i} \eta_{i,a} e^{-\beta (U_{i} - V \sum_{b} \eta_{i,b}\sigma_{b})}
\end{equation}
where $Z$ is the partition function
\begin{equation}
    Z = \sum_{i} e^{-\beta (U_{i} - V \sum_{b} \eta_{i,b}\sigma_{b})} \; .
\end{equation}
Here $i$ runs over microscopic states of the simulated system,
$\beta=(k_B T)^{-1}$ where $k_B$ is Boltzmann's constant, $V$ is
the equilibrium volume of the simulation supercell at the considered
conditions of temperature and stress, and $U_i$ and
$\boldsymbol{\eta}_i$ are, respectively, the energy and strain of
state $i$. Then, by taking the derivative of $\langle \eta_a \rangle$
with respect to the applied stress, we can readily obtain
\begin{equation}
    S_{ab}=\beta V [\langle \eta_{a} \eta_{b} \rangle - \langle \eta_{a} \rangle \langle \eta_{b} \rangle] \; ,
\end{equation}
which is our linear-response formula for the compliance. As usually
done in simulation studies of piezoelectric properties that involve
similar quantities~\cite{garcia98}, here we neglect the stress
derivative of the volume; we explicitly checked this is an accurate
approximation.

To finish this section, let us note that in this study we define
strains by taking as a reference the lattice constant of the cubic
phase of PbTiO$_{3}$, as obtained from a symmetry-constrained
first-principles simulation at 0~K. This reference value is $l_0
=$~3.88~\AA. Hence, for the computed normal strains $\eta_{a}$ (where
$a = 1, 2, 3$ labels the pseudo-cubic perovskite axes), the
corresponding lattice constants are given by
\begin{equation}
    l_{a} = l_{0}(1+\eta_{a}) \; .
\end{equation}
Note that we can use this simple expression because shear strains
($\eta_{a}$ with $a=4,5,6$) are zero at all temperatures in
PbTiO$_{3}$.

Finally, in the following we choose the Cartesian axes to coincide
with the pseudo-cubic directions of the perovskite lattice. Also, we
assume that the tetragonal ferroelectric phase of PbTiO$_{3}$ is
characterized by a positive polarization along the third Cartesian
axis.

\section{Results and discussion}

We now present our results, emphasizing and discussing the most
interesting aspects revealed by our calculations. We also compare with
available experimental information.

\begin{figure}[t!]
\includegraphics[width=1.0\linewidth]{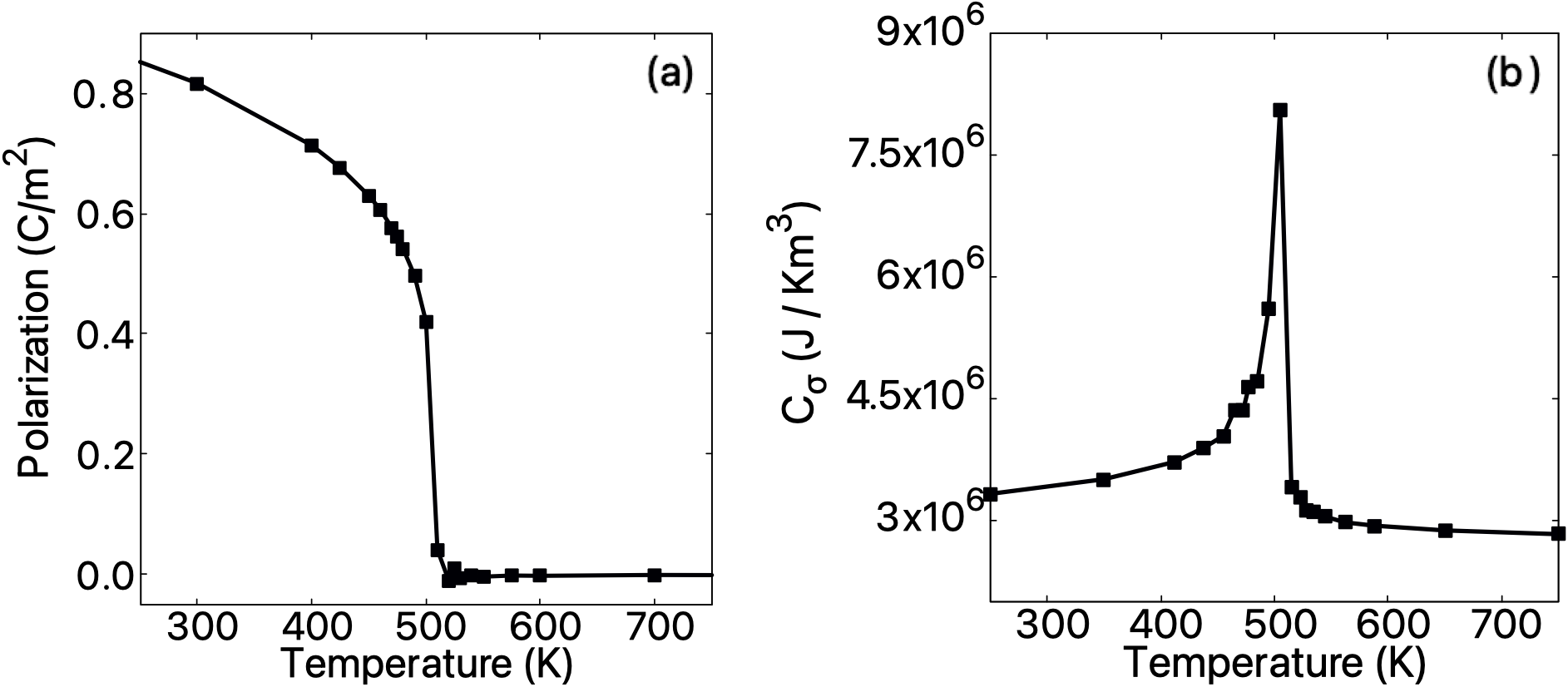}
\caption{Computed polarization ${\bf P}=(0, 0, P)$ (a) and specific heat $C_{\sigma}$ (b) as a function of temperature.}
\label{fig:basic-mc} 
\end{figure}

\subsection{Basic Monte Carlo results}

In Fig.~\ref{fig:basic-mc} we present the temperature dependence of
the polarization as obtained from our Monte Carlo simulations
(panel~(a)) as well as our results for the specific heat
(panel~(b)). We obtain a weakly first-order ferroelectric phase
transition at $T_{\rm C}=$~510~K, where one polarization component
becomes different from zero in a discontinuous fashion. This
transition, from the high-temperature cubic ($Pm\bar{3}m$) phase to a
tetragonal polar ($P4mm$) state, agrees with the well-known
experimental behavior of PbTiO$_{3}$ \cite{lines-book1977} (albeit the
underestimated transition temperature mentioned above) and with
previous simulations using the same second-principles
potential~\cite{wojdel13}. The specific heat presents an anomaly (a
maximum) at the transition temperature, also as expected.

\begin{figure*}[t!]
\includegraphics[width=0.95\linewidth, height=0.385\textheight]{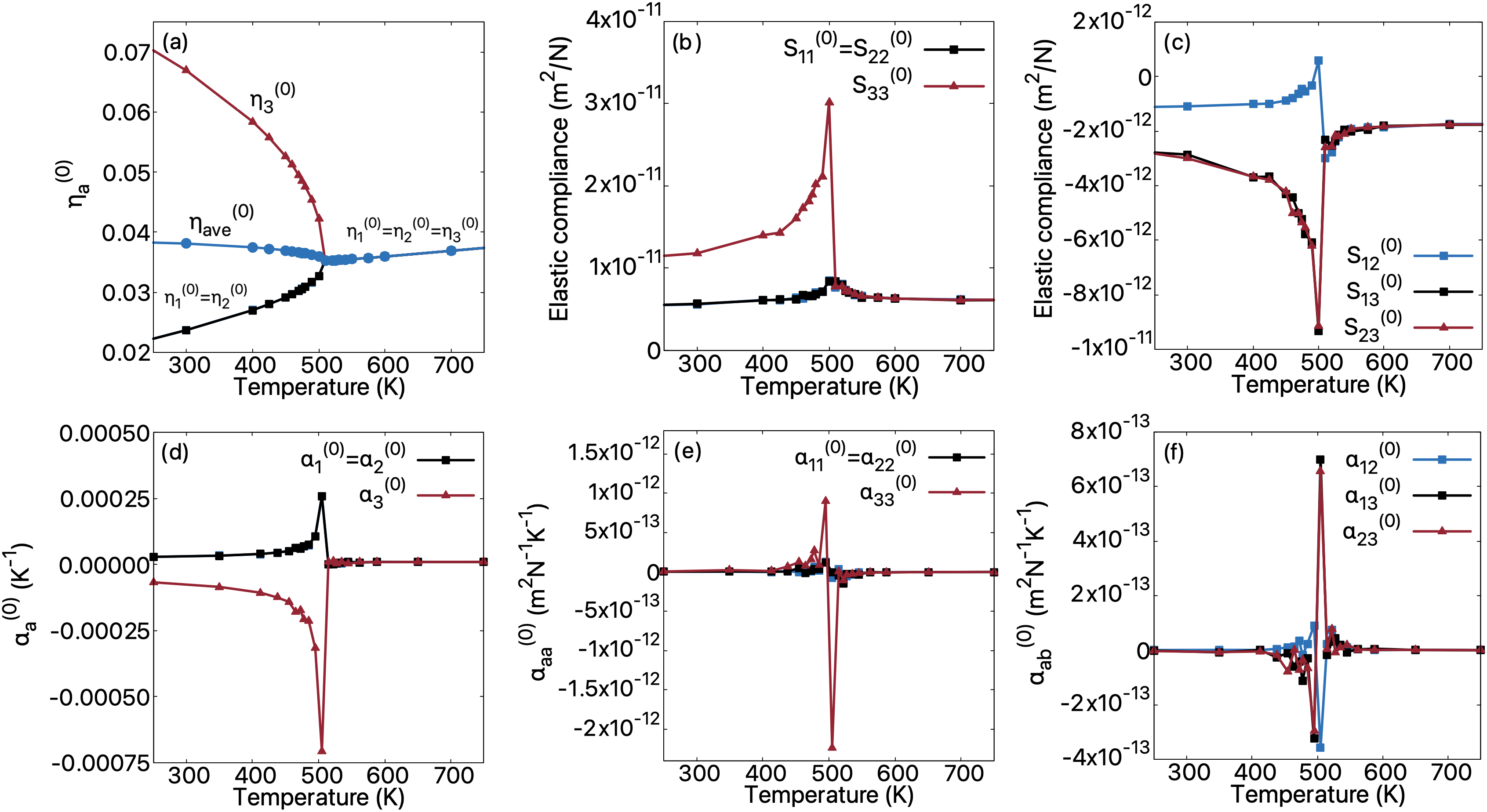}
\caption{Computed temperature dependence of key strain-related quantities. We present the results for the normal strains (a), including the average value $\eta_{\rm ave}$; the diagonal normal components of the compliance tensor (b); the off-diagonal normal components of the compliance tensor (c); the $T$-derivatives of the strains (d); the $T$-derivatives of the diagonal normal components of the compliance tensor (e); and the $T$-derivatives of the off-diagonal normal components of the compliance tensor (e).}
\label{fig:elastic} 
\end{figure*}

In Fig.~\ref{fig:elastic} we present how the strain-related quantities
evolve across this phase transition. Panel~(a) shows the temperature
evolution of the normal strains ($\eta_{1}$, $\eta_{2}$ and
$\eta_{3}$) as well as their average ($\eta_{\rm ave}$, which accounts
-- to first order -- for variations in volume). (We do not show the
shear strains, which are zero at all temperatures.) The symmetry
breaking associated to the ferroelectric transition is clearly visible
in the figure. Indeed, this transformation has an {\em improper}
ferroelastic character, whereby we observe strain changes following
the onset of the spontaneous polarization.

More precisely, Fig.~\ref{fig:elastic}(a) shows two differentiated
behaviors. In the paraelectric phase, for $T>T_{\rm C}$, the three
strain components are equal ($\eta_1=\eta_2=\eta_3$) and increase
gradually with temperature. Hence, we have an isotropic positive
thermal expansion. In contrast, when we heat up the material from low
temperatures in the ferroelectric phase ($T<T_{\rm C}$), the strain
component parallel to the polarization ($\eta_{3}$) decreases with
increasing temperature, while the components perpendicular to it
($\eta_{1}=\eta_{2}$) grow. This reflects the presence of a tetragonal
distortion (usually quantified by the $c/a$ aspect ratio of the
tetragonal unit cell), which can be shown to be proportional to the
square of the spontaneous polarization (see, e.g., the discussion in
Ref.~\onlinecite{kingsmith94}). Interestingly, the combination of
these normal strains results in an average deformation that decreases
with increasing temperature, yielding a negative thermal expansion for
$T<T_{\rm C}$. This effect has been discussed in the literature, e.g.,
in Refs.~\onlinecite{ritz19} and \onlinecite{chen15}.

Figures~\ref{fig:elastic}(b) and \ref{fig:elastic}(c) show our results
for the components of the elastic compliance tensor involving normal
strains. The symmetry breaking from cubic ($S_{11}=S_{22}=S_{33}$,
$S_{12}=S_{13}=S_{23}$) to tetragonal ($S_{11}=S_{22}\neq S_{33}$,
$S_{12}\neq S_{13}=S_{23}$) is obvious here as well. We also find
that, in the ferroelectric phase, the direction of the polarization is
elastically softer than the perpendicular plane
($S_{33}>S_{11}=S_{22}$). Further, all compliance components reach
maximum values around the transition point, presenting a simple
monotonic variation at all other temperatures.

Finally, in panels (d) to (f) of Fig~\ref{fig:elastic} we show the
temperature derivatives of the previous quantities, which we obtain
numerically from the data in the first three panels. These are the
$\alpha_{a}^{(0)}$ and $\alpha^{(0)}_{ab}$ tensor components
introduced above, the key ingredients to compute mechanocaloric
temperature changes within our perturbative formalism.

\begin{figure*}[t!]
\includegraphics[width=0.95\linewidth, height=0.2\textheight]{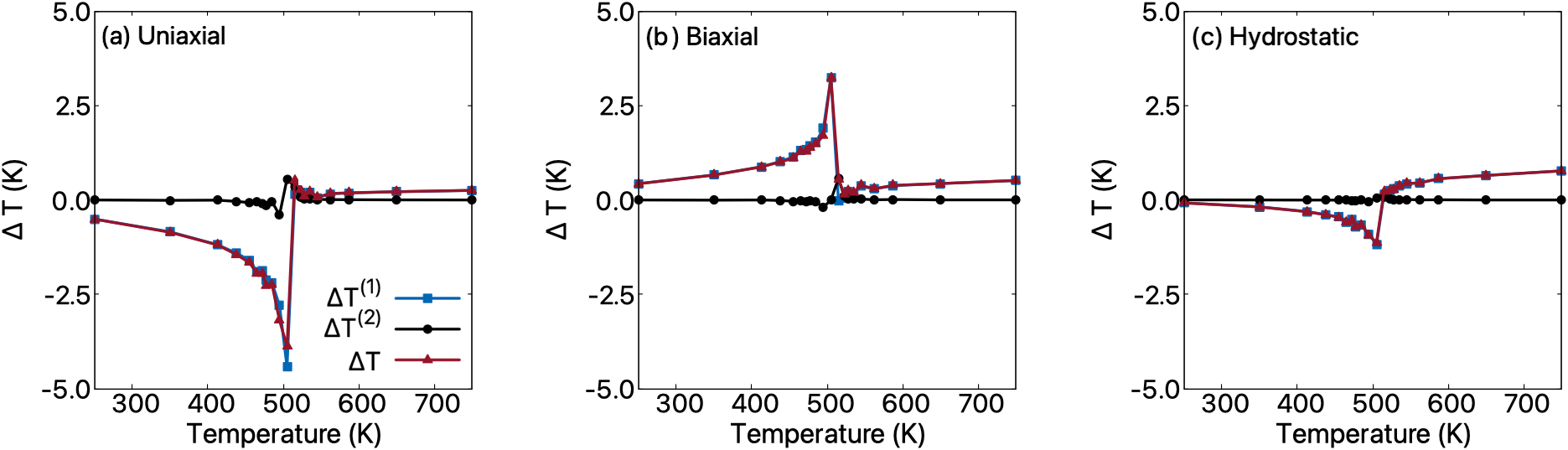}
\caption{Mechanocaloric effect as a function of temperature dependence
  for a compressive stress of $-0.1$~GPa applied (a) along the
  polarization direction, (b) in the plane normal to \textbf{P} and
  (c) hydrostatically.}
\label{fig:deltaT} 
\end{figure*}

\subsection{Mechanocaloric temperature change}

The results in Figs.~\ref{fig:basic-mc}(b) and \ref{fig:elastic} allow
us to compute the adiabatic mechanocaloric temperature change using
the formalism introduced above. Representative results are given in
Fig.~\ref{fig:deltaT}, all corresponding to the application of a
compressive stress of $-$0.1~GPa. We consider a uniaxial stress along
the polarization direction in panel~(a), a biaxial stress in the plane
perpendicular to ${\bf P}$ in panel~(b), and a hydrostatic pressure in
panel~(c). In all cases we show the total $\Delta T$ as well as the
individual contributions with a linear ($\Delta T^{(1)}$) and
quadratic ($\Delta T^{(2)}$) dependence on the stress.

\begin{figure}[b!]
\includegraphics[width=1.0\linewidth]{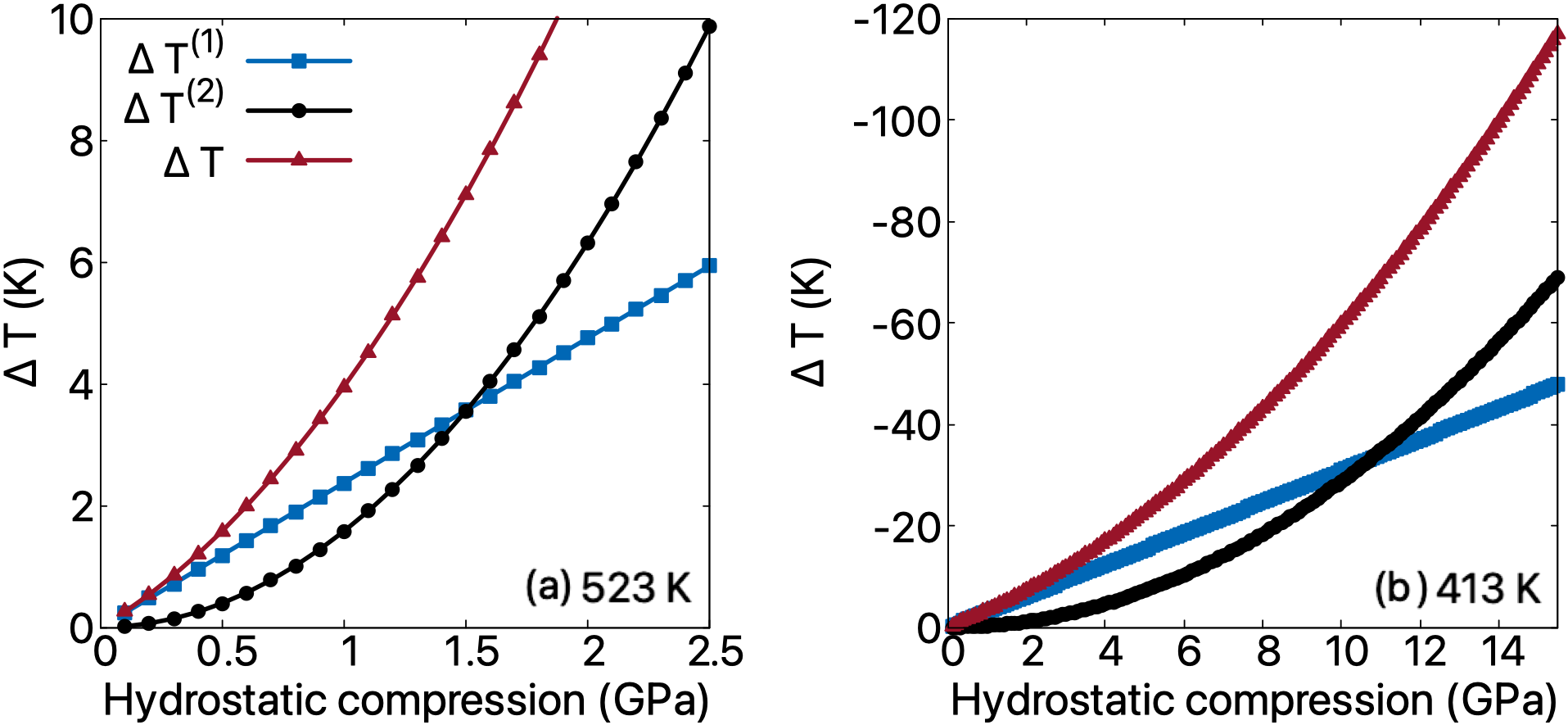}
\caption{Linear and quadratic contributions to the barocaloric effect
  as a function of applied compression at 523~K (a) and 413~K (b).}
\label{fig:deltaT_vs_sigma} 
\end{figure}

Two aspects are common to all the results in
Fig.~\ref{fig:deltaT}. First, the largest temperature changes are
obtained in the vicinity of the phase transition. This is consistent
with the results in Fig.~\ref{fig:elastic}, which shows that all the
relevant quantities ($T$-derivatives) present their largest absolute
values around $T_{\rm C}$. Second, for the considered $-$0.1~GPa, the
linear effect dominates the caloric response, even in the vicinity of
the phase transition. As a matter of fact, the linear approximation to
$\Delta T$ is a good one in a wide range of applied stress; indeed, as
shown in Fig.~\ref{fig:deltaT_vs_sigma}, we need to reach compressions
of about $-$1.5~GPa for the quadratic effect to dominate at
temperatures close to $T_{\rm C}$ (see results for 523~K in
panel~(a)), and even higher when we move away from the transition
temperature ($-$10.7~GPa at $T=413$~K, as shown in panel~(b)).

We also find that, in all cases, the obtained $\Delta T$ is positive
and relatively small above $T_{\rm C}$, as consistent with the
computed small isotropic thermal expansion at high temperatures (see
Fig.~\ref{fig:elastic}(a)). A marked anisotropy of the caloric responses
appears when we move below $T_{\rm C}$, reflecting the lower symmetry
of the tetragonal ferroelectric state.

Indeed, most importantly, we find that the ferroelectric state may
present a conventional ($\Delta T>0$) or inverse ($\Delta T < 0$)
effect depending on how the compressive stress is applied. When we
compress in the plane perpendicular to the polarization
(Fig.~\ref{fig:deltaT}(b)), we obtain $\Delta T>0$ reflecting the
positive thermal expansion of the $a$ and $b$ lattice
constants. In contrast, a compression along the polar axis
(Fig.~\ref{fig:deltaT}(a)) yields a negative temperature change, as a
result of the reduction of the $c$ lattice constant upon heating.

Hence we obtain a remarkable result: because of the development of the
ferroelectric polarization and its impact in the lattice strains,
PbTiO$_{3}$ turns out to be an extremely anisotropic mechanocaloric,
to the point that the sign of the temperature change depends on the
direction of the applied stress. As far as we know, this is the first
example of mechanocaloric material presenting such a {\em sign
  anisotropy}. (Examples of materials that can switch between
conventional and inverse behaviors are known,
though~\cite{alvarezalonso17,oadira20,xiao21}.)

Note also that the relationship just described between polarization
and strain is not exclusive of PbTiO$_{3}$, but typical of
ferroelectric perovskites (see, e.g., the behavior of the tetragonal
phase of BaTiO$_{3}$~\cite{megaw47,jona-book1993}) and other
ferroelectric families (see, e.g., the case of
LiTaO$_{3}$~\cite{bartasyte12}). Hence, the coexistence of
conventional and inverse mechanocaloric responses may be a common
feature among ferroelectrics. In addition, materials with a negative
thermal expansion (where, often, some lattice constants decrease and
some increase upon heating~\cite{goodwin08,collings16,chen15}) are
also good candidates to present such a behavior.

It is also worth to mention that our computed temperature changes for
uniaxial and biaxial compression are remarkably large, despite the
modest stress of $-$0.1~GPa here considered. Indeed, as shown in
Figs.~\ref{fig:deltaT}(a) and \ref{fig:deltaT}(b), we get temperature
changes exceeding 3~K close to the phase transition (510~K), and still
notable as we move into the ferroelectric state (e.g., above 1~K for
all temperatures between 450~K and 510~K). Also remarkably, the gap
between the positive and negative effects is as large as 7~K at
$T_{\rm C}$.

In contrast, the response to a hydrostatic pressure
(Fig.~\ref{fig:deltaT}(c)) is relatively small within the
ferroelectric phase, as it suffers from the partial cancellation of
the conventional and inverse effects just discussed. We have a maximum
barocaloric $\Delta T \approx -1$~K at $T_{\rm C}$, the inverse effect
being dominant. This is a direct consequence of the negative
volumetric thermal expansion obtained for $T<T_{\rm C}$ and shown in
Fig.~\ref{fig:elastic}(a). Naturally, as we cross $T_{\rm C}$ and the
isotropic thermal expansion changes from negative to positive, so does
the sign of the barocaloric response. The same applies to the
elastocaloric response to an uniaxial stress along the polar axis.

\begin{figure*}[ht!]
\includegraphics[width=1\linewidth, height=0.21\textheight]{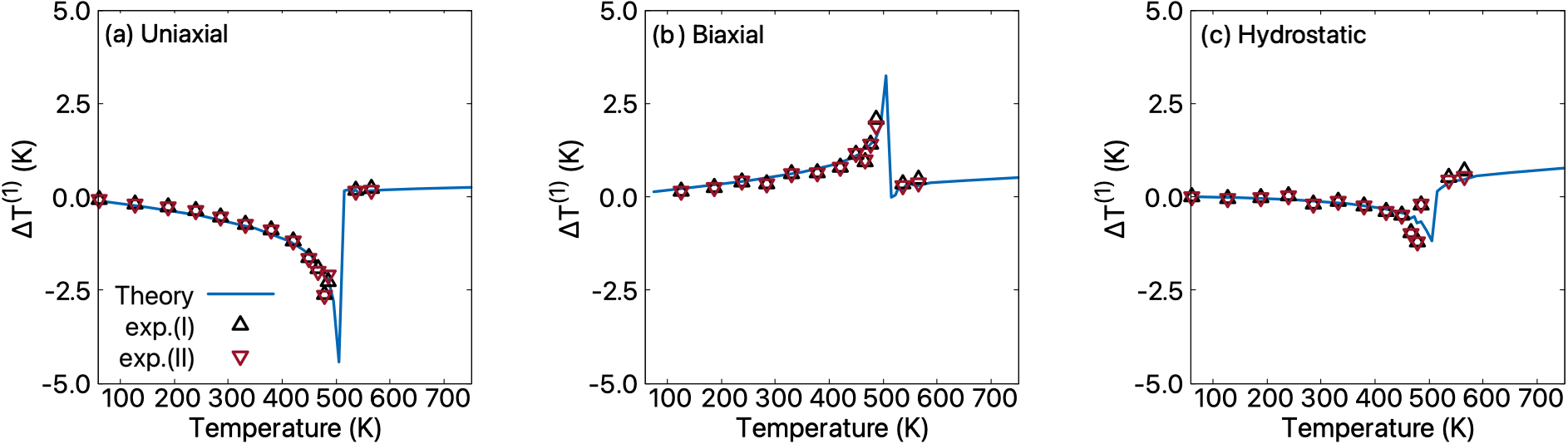}
\caption{Comparison between the theoretical prediction and two
  experimental-based estimates of the linear contribution to the
  mechanocaloric response, for a compressive stress of $-$0.1~GPa
  applied along the polarization direction (a), in the plane normal to
  \textbf{P} (b) and hydrostatically (c). As explained in the text,
  ``exp.~(I)'' labels the case where we combine the experimental
  thermal expansion with the theoretical specific heat, while
  ``exp.~(II)'' corresponds to the case in which both thermal
  expansion and specific heat are taken from experiment.}
\label{fig:comp_with_exp} 
\end{figure*}

\subsection{Comparison with previous works}

We can compare our theoretical predictions with experiment in
different ways.

First, based on experimental results for the temperature dependence of
the lattice constants of PbTiO$_{3}$ (we used those reported in
Ref.~\cite{haun87}), we can numerically obtain the spontaneous thermal
expansion vector $\boldsymbol{\alpha}^{(0)}$. Then, we shift down the
experimental temperatures to make the transition point coincide with
the calculated $T_{\rm C} = 510$~K. With this data, combined with our
calculated results for $C_{\sigma}$, we can trivially estimate $\Delta
T^{(1)}$, i.e., the part of the mechanocaloric response that is linear
in the applied stress and which, according to our calculations,
dominates the effect for moderate
compressions. Figure~\ref{fig:comp_with_exp} shows the results for
$\Delta T^{(1)}$ thus obtained (symbols labeled ``exp.~(I)''),
together with our theoretical predictions. We find that the agreement
between theory and the experiment-based estimate is essentially
perfect, indicating that our simulations capture very well (save the
error in $T_{\rm C}$) the thermal evolution of PbTiO$_{3}$'s
lattice. In particular, the change of sign of the barocaloric effect
across $T_{\rm C}$ is readily recovered when using the experimental
data for $\boldsymbol{\alpha}^{(0)}$, which displays the same reversal
across the Curie point (from negative to positive thermal expansion)
as computed.

Second, we deduce $\Delta T^{(1)}$ exclusively from experimental
information, by combining the structural results of
Ref.~\onlinecite{haun87} with the calorimetric data of
Ref.~\onlinecite{mikhaleva12} (shifted too, so that the experimental
and theoretical $T_{\rm C}$'s coincide). We thus obtain the symbols
labeled ``exp.~(II)'' in Fig.~\ref{fig:comp_with_exp}, which are in
excellent agreement with our theoretical values and our first
experimental estimate (``exp.~(I)''). This indicates that the
(temperature-shifted) specific heat is in excellent agreement with our
predictions.

Indeed, for example, Refs.~\onlinecite{mikhaleva12} and
\onlinecite{rossetti2005} report values between
2.6~MJ~m$^{-3}$~K$^{-1}$ and 2.9~MJ~m$^{-3}$~K$^{-1}$ for $C_{\sigma}$
near room temperature, while our computed value at 300~K is about
3.4~MJ~m$^{-3}$~K$^{-1}$. Further, after the mentioned temperature
shift, we obtain $C_{\sigma} = 3.4$~MJ~m$^{-3}$~K$^{-1}$ at 300~K from
the data in Ref.~\onlinecite{mikhaleva12}, in perfect agreement with
our calculation. This suggests that our computational approach
predicts PbTiO$_{3}$'s thermal properties with remarkable accuracy.

Finally, the barocaloric effect in PbTiO$_{3}$ was reported in
Ref.~\onlinecite{mikhaleva12}, which in principle allows us to make a
direct comparison. Below $T_{\rm C}$, this article describes an
inverse behavior that is similar to the one we predict, with
quantitative results (e.g., $\Delta T \approx -2$~K at the transition
temperature for a pressure of $-$0.03~GPa) in reasonable agreement
with our computed values. However, Ref.~\onlinecite{mikhaleva12} also
reports that the negative temperature change persists above $T_{\rm
  C}$ for all the measured pressures (e.g., $\Delta T \approx -0.1$~K
is measured at about 25~K degrees above $T_{\rm C}$ for $-$0.03~GPa),
which would suggest that PbTiO$_{3}$ presents a negative thermal
expansion in its paraelectric phase. This is in direct contradiction
with abundant experimental \cite{bhide68, shirane56, shirane50} and
theoretical \cite{wojdel13, sepliarsky04, nishimatsu12} studies of the
structural evolution of this compound. Hence, we are not sure about
the status of the barocaloric results of Ref.~\onlinecite{mikhaleva12}
or to what extent we should expect agreement with our predictions.

Additionally, we should stress we have not found any experimental
investigation addressing the (strong) anisotropy of the mechanocaloric
response in the polar phase of PbTiO$_{3}$. This question, which
constitutes one of our most interesting predictions, remains to be
explicitly verified experimentally.

Finally, let us note that our results are consistent with other
theoretical studies of PbTiO$_3$. For instance, using a
first-principles-based effective Hamiltonian, Lisenkov {\sl et al.}
\cite{lisenkov13} predict $\Delta T = +6$~K under the application of a
tensile stress of $+$0.2~GPa along the polar direction and near the
transition temperature; this is in acceptable agreement with our
computed $\Delta T \approx -4$~K upon an uniaxial compression of
$-$0.1~GPa close to $T_{\rm C}$. Our results are also consistent the
effective-Hamiltonian study of Barr {\sl et al.}~\cite{barr15}, who
report a temperature change of about $-5$~K upon {\em release} of a
tensile load of $+0.2$~GPa close to $T_{\rm C}$. Similarly, a
phenomenological study of PbTiO$_{3}$ and related ferroelectrics
\cite{khassaf17} reports temperature increases of up to $+$2~K for a
tensile uniaxial pressure of $+$0.1~GPa along the polarization
direction and close to $T_{\rm C}$. Hence, all these literature
results are quantitatively and qualitatively consistent with the
dominant linear effect revealed in the present work, whereby the sign
of $\Delta T \approx \Delta T^{(1)}$ depends on the nature (tensile or
compressive) of the applied stress.

\section{Summary and conclusions}

In this work we have used predictive atomistic simulation
(``second-principles'') methods to investigate the intrinsic
mechanocaloric response of prototype ferroelectric
PbTiO$_{3}$. Notably, we find that the effects can be quite large in
the vicinity of the ferroelectric Curie point (up to $-$4~K for
compressions of only $-$0.1~GPa), even though no phase transition is
induced in the material (hence, our obtained values do not have any
latent-heat contribution).

Remarkably, we reveal that the mechanocaloric response is strongly
anisotropic in the ferroelectric phase of the compound. More
precisely, we find that the effect is conventional (temperature
increases under compression) if a stress is applied in the plane
perpendicular to the spontaneous polarization, and inverse
(temperature decreases) if we compress along the polarization
direction. As far as we know, such a coexistence of conventional and
inverse responses had never been observed or predicted before in any
material, and remains to be explicitly confirmed
experimentally. Nevertheless, we find that our results are compatible
with available experimental information on the elastic properties of
PbTiO$_{3}$, which suggests that the predicted coexistence is real.

Our theoretical calculations rely on a perturbative formalism that we
introduce here and which should be useful in the broader context of
mechanocaloric studies, including experimental ones. (We illustrate
this explicitly when checking our predictions against experimental
information.) We should emphasize that this perturbative theory
applies whenever the external stress is not as large as to induce a
discontinuous phase change in the material. Hence, the caloric
response it captures is eminently reversible and does not include any
latent-heat contribution. While, admittedly, the best mechanocaloric
materials largely base their performance on the latent heat released
(absorbed) at stress-driven first-order phase
transitions~\cite{engelbrecht19,kabirifar19,imran21,chauhan15}, the
present scheme allows us to inspect in detail the behavior within the
range of continuous deformations, and is the key to the most
interesting conclusions of this work. For example, thanks to this
perturbative formalism, we can determine the dominance of the
lowest-order temperature change (linear in the applied stress) for the
moderate compressions used here (typically, $-$0.1~GPa), and thus
predict (and explain) the coexistence of conventional and inverse
mechanocaloric responses in the ferroelectric state.

Along the same lines, our perturbative theory plainly shows that the
key to the predicted coexistence of conventional and inverse effects
lies in the differentiated temperature dependence of the strains in
the polar phase of PbTiO$_{3}$: the strain parallel to the spontaneous
polarization decreases upon heating, while the perpendicular strains
increase. Interestingly, this feature is shared by many
ferroelectrics~\cite{megaw47,jona-book1993,bartasyte12}; and a similar
anisotropic behavior of the strains is typical of compounds exhibiting
a negative thermal expansion~\cite{goodwin08,collings16,chen15}. Thus,
our analysis suggests that such compounds are likely to present a
coexistence of conventional and inverse mechanocaloric responses.

We hope this work will encourage further investigations of the
(potentially exotic and large) mechanocaloric properties of materials
seldom considered to that end, such as ferroelectrics. We also hope
that the simple perturbative formulas introduced here will be useful
in future mechanocaloric studies, both theoretical and experimental.

Work funded by the Luxembourg National Research Fund (FNR) through
grant FNR/C18/MS/12705883/REFOX/Gonzalez. Additionally, M.G. was
supported by FNR Grant INTER/RCUK/18/12601980.

\end{document}